\documentclass[nofootinbib,aps,11pt]{revtex4}
\usepackage{graphicx}
\usepackage{cancel}
\usepackage{amssymb}
\usepackage{textcomp}
\usepackage{amsmath}
\usepackage{bm}
\usepackage{times}
\usepackage{epsfig}
\usepackage{color}
\usepackage{axodraw}

\begin{document}

\title{\Large Explaining Solar Neutrinos 
with Heavy Higgs Masses in
Partial Split Supersymmetry}
\author{Marco Aurelio D\'\i az}
\author{Francisca Garay}
\author{Benjamin Koch}
\affiliation{
{\small Departamento de F\'\i sica, Pontificia Universidad Cat\'olica 
de Chile, Avenida Vicu\~na Mackenna 4860, Santiago, Chile 
}}
%%%%%%%%%%%%%%%%%%%%%%%%%%%%%%%%%%%%%%%%%%%%%%%%%%%%%%%%%%%%%%%%%%%%
\begin{abstract}
Partial Split Supersymmetry with violation of R-parity as a model
for neutrino masses is explored. It is shown that
at the one-loop level the model can give predictions that
are in agreement with all present experimental values
for the neutrino sector. An analytical result is that the
small solar neutrino mass difference can be naturally
explained in the decoupling
limit for the heavy Higgs mass eigenstates.
\end{abstract}
%%%%%%%%%%%%%%%%%%%%%%%%%%%%%%%%%%%%%%%%%%%%%%%%%%%%%%%%%%%%%%%%%%%%
\date{\today}
\maketitle
%%%%%%%%%%%%%%%%%%%%%%%%%%%%%%%%%%%%%%%%%%%%%%%%%%%%%%%%%%%%%%%%%%%%
\section{Introduction}
%%%%%%%%%%%%%%%%%%%%%%%%%%%%%%%%%%%%%%%%%%%%%%%%%%%%%%%%%%%%%%%%%%%

Supersymmetric models softly broken at low energies introduce new 
sources of flavor changing neutral currents and new CP-violating 
phases that can create un-observed phenomenological effects. If 
supersymmetric particles are of the order of the electroweak scale, 
it is not obvious why they do not contribute to processes with flavor changing
neutral currents
with rates higher than observed \cite{Nir:2007xn}. Large values 
of sparticle masses is one mechanism that explains the inconspicuous 
contributions from supersymmetry to flavor observables. Split
supersymmetric models were introduced advertising precisely this 
feature \cite{ArkaniHamed:2004fb}.

R-Parity violation in supersymmetric models \cite{Barbier:2004ez}
is an attractive feature because it provides a mechanism for neutrino 
mass generation. It is specially compelling in the case of Bilinear 
R-Parity violation (BRpV) \cite{Hempfling:1995wj} because an 
atmospheric mass difference is generated at tree-level by a low 
energy see-saw mechanism triggered by a mixing between neutrinos and 
neutralinos. In addition, a solar mass difference is generated at 
one-loop level by contributions from all particles 
\cite{Hirsch:2000ef}, explaining the hierarchy between atmospheric and 
solar mass scales. Original Split Supersymmetry (SS) conserves 
R-Parity, nonetheless, the possibility has been explored, proving for
example that a SS with R-Parity violation cannot generate a solar 
mass \cite{Chun:2004mu,Diaz:2006ee,Davidson:2000ne,Davidson:2000uc}.

In Partial Split Supersymmetry (PSS) \cite{Diaz:2006ee}, all sfermions 
are very heavy and decouple from the low energy theory, alleviating the 
flavor constraints present in supersymmetric models. Nevertheless, as 
opposed to the original Split Supersymmetric model, both Higgs boson 
doublets remain light in comparison with the split supersymmetric scale 
$\widetilde m$. Additionally we assume that R-Parity is not conserved. As a
consequence, 
a solar neutrino mass difference is generated at one-loop, while the 
atmospheric mass difference is generated at tree level.

In contrast to gravitationally inspired models \cite{Diaz:2009yz},
in PSS the solar neutrino mass difference is generated by loops 
involving the CP-odd Higgs $A$ and the CP-even Higgs bosons $h$ and $H$. We 
are particularly interested in the limit $m_A\gg m_Z$, called the 
decoupling limit \cite{Gunion:2002zf}, where the light Higgs $h$ 
decouples from the heavy ones, acquiring SM-like couplings to fermions. 
Under this condition, the Higgs boson $h$ does not contribute to the 
solar mass, and it is experimentally challenging to distinguish it from 
the SM Higgs boson. In this article we study this scenario, and the 
ability of the mechanism to generate a solar mass from loops involving 
the heavy Higgs bosons $H$ and $A$.

%%%%%%%%%%%%%%%%%%%%%%%%%%%%%%%%%%%%%%%%%%%%%%%%%%%%%%%%%%%%%%%%%%
\section{R-parity Violation and Neutrino Masses in PARTIAL 
SPLIT SUSY}
%%%%%%%%%%%%%%%%%%%%%%%%%%%%%%%%%%%%%%%%%%%%%%%%%%%%%%%%%%%%%%%%%%

Partial Split Supersymmetry is an effective theory whose lagrangian is 
valid at scales lower than $\widetilde m$. The low energy lagrangian 
includes two Higgs doublet $H_u$ and $H_d$, and it is characterized by 
the following R-Parity conserving terms,
\begin{eqnarray}
{\cal L}_{PSS}^{RpC} &\owns& - \Big[ m_1^2H_d^\dagger H_d 
+ m_2^2H_u^\dagger H_u - m_{12}^2(H_d^T\epsilon H_u+h.c.)
\nonumber\\ &&
+ \textstyle{1\over2}\lambda_1(H_d^\dagger H_d)^2 
+ \textstyle{1\over2}\lambda_2(H_u^\dagger H_u)^2 
+ \lambda_3(H_d^\dagger H_d)(H_u^\dagger H_u)
+ \lambda_4|H_d^T\epsilon H_u|^2
\Big]
\nonumber\\ &&
+ h_u \overline u_R H_u^T \epsilon q_L 
- h_d \overline d_R H_d^T \epsilon q_L
- h_e \overline e_R H_d^T \epsilon l_L \ - \
\label{LagSS2HDM}\\ &&
-\textstyle{\frac{1}{\sqrt{2}}} H_u^\dagger
(\tilde g_u \sigma\widetilde W + \tilde g'_u\widetilde B)\widetilde H_u
-\textstyle{\frac{1}{\sqrt{2}}} H_d^\dagger
(\tilde g_d \sigma \widetilde W - \tilde g'_d \widetilde B)\widetilde H_d 
+\mathrm{h.c.}
\nonumber
\end{eqnarray}
In the first two lines we have the Higgs potential, including both mass 
and self interaction terms. The electroweak symmetry is spontaneously broken
as in the MSSM when the Higgs fields acquire non vanishing vacuum expectation
values.  In the third line we have the Yukawa 
interactions, and in the fourth line we have the interactions between
Higgs, gauginos and higgsinos. This lagrangian is to be compared with 
the supersymmetric lagrangian valid above the scale $\widetilde m$, which
includes the analogous terms,
\begin{eqnarray}
{\cal L}_{susy}^{RpC} &\owns& - \Big[ m_1^2H_d^\dagger H_d 
+ m_2^2H_u^\dagger H_u - m_{12}^2(H_d^T\epsilon H_u+h.c.)
+ \textstyle{1\over8}(g^2+g'^2)(H_d^\dagger H_d)^2
\nonumber\\ &&
+ \textstyle{1\over8}(g^2+g'^2)(H_u^\dagger H_u)^2
+ \textstyle{1\over4}(g^2-g'^2)(H_d^\dagger H_d)(H_u^\dagger H_u)
- \textstyle{1\over2}g^2|H_d^T\epsilon H_u|^2
\Big]
\nonumber\\ &&
+ \lambda_u \overline u_R H_u^T \epsilon q_L 
- \lambda_d \overline d_R H_d^T \epsilon q_L
- \lambda_e \overline e_R H_d^T \epsilon l_L
\label{LagSplit2}\\ &&
-\textstyle{\frac{1}{\sqrt{2}}} H_u^\dagger
(g \sigma\widetilde W + g'\widetilde B)\widetilde H_u
-\textstyle{\frac{1}{\sqrt{2}}} H_d^\dagger
(g \sigma \widetilde W - g' \widetilde B)\widetilde H_d 
+\mathrm{h.c.}
\nonumber
\end{eqnarray}
These two models are connected through boundary conditions at the scale 
$\widetilde m$. For the Higgs self couplings they are,
\begin{equation}
\lambda_1=\lambda_2=\textstyle{1\over4}(g^2+g'^2) \,, \qquad
\lambda_3=\textstyle{1\over4}(g^2-g'^2) \,, \qquad
\lambda_4=-\textstyle{1\over2}g^2,
\end{equation}
which are typical matching conditions between the MSSM and two Higgs 
doublet models. In an analogous way we have for the Yukawa couplings at 
$\widetilde m$,
\begin{equation}
h_u=\lambda_u\,,\qquad h_d=\lambda_d\,,\qquad 
h_e=\lambda_e\,, 
\end{equation}
and for the higgsino-gaugino Yukawa couplings at $\widetilde m$, 
\begin{equation}
\tilde g_u=\tilde g_d=g\,,\qquad
\tilde g'_u=\tilde g'_d=g'\,. 
\end{equation}
The two Higgs fields $H_u$ and $H_d$ acquire a vacuum expectation value 
$v_u$ and $v_d$, defining the usual mixing angle $\tan\beta=v_u/v_d$.
The neutralino mass matrix in this scenario is written as,
\begin{equation}
{\bf M}_{\chi^0}^{PSS}=\left[\begin{array}{cccc}
M_1 & 0 & -\frac{1}{2}\tilde g'_d v_d & \frac{1}{2}\tilde g'_u v_u \\
0 & M_2 & \frac{1}{2}\tilde g_d v_d & -\frac{1}{2}\tilde g_u v_u \\
-\frac{1}{2}\tilde g'_d v_d & \frac{1}{2}\tilde g_d v_d & 0 & -\mu \\
\frac{1}{2}\tilde g'_u v_u & -\frac{1}{2}\tilde g_u v_u & -\mu & 0
\end{array}\right].
\label{X0massmat2}
\end{equation}
In Partial Split Supersymmetry with non conserved R-Parity,
neutrino masses are generated \cite{Diaz:2006ee}. Trilinear RpV terms are
irrelevant because high sfermions masses make their loop contributions 
negligible. Bilinear RpV terms do contribute via neutrino/neutralino mixing. 
The relevant terms are,
\begin{equation}
{\cal L}_{PSS}^{RpV} =
-\epsilon_i \widetilde H_u^T \epsilon L_i  
\ -\ 
\textstyle{\frac{1}{\sqrt{2}}} b_i H_u^T\epsilon
(\tilde g_d \sigma\widetilde W-\tilde g'_d\widetilde B)L_i 
\ + \ h.c., 
\label{LSS2HDMRpV}
\end{equation}
where $\epsilon_i$ are the usual mass parameters that mix Higgs with lepton
superfields, and $b_i$ are effective couplings between Higgs, gauginos and 
leptons. After the Higgs fields acquire vacuum expectation values, mixing 
terms are generated between neutrinos, on one hand, and higgsinos and 
gauginos on the other hand,
\begin{equation}
{\cal L}^{RpV}_{PSS} = - \left[
\epsilon_i \widetilde H_u^0 + \frac{1}{2} b_i v_u \left( 
\tilde g_d \widetilde W_3 - \tilde g'_d \widetilde B \right)
\right] \nu_i \ + \ h.c. \ + \ \ldots
\end{equation}
and they extend the $4\times4$ neutralino mass matrix, in 
eq.~(\ref{X0massmat2}), to a $7\times7$ mass matrix that includes the 
neutrinos. The off-diagonal mixing block is,
\begin{equation}
m^{PSS}=\left[\begin{array}{cccc}
-\frac{1}{2} \tilde g'_d b_1 v_u & 
 \frac{1}{2} \tilde g_d  b_1 v_u & 
0 &\epsilon_1 
\cr
-\frac{1}{2} \tilde g'_d b_2 v_u & 
 \frac{1}{2} \tilde g_d  b_2 v_u & 
0 & \epsilon_2 
\cr
-\frac{1}{2} \tilde g'_d b_3 v_u & 
 \frac{1}{2} \tilde g_d  b_3 v_u &
0 & \epsilon_3
\end{array}\right],
\end{equation}
while the neutrino-neutrino block is a null $3\times3$ matrix. After a 
low energy see-saw mechanism, the effective neutrino mass matrix is,
\begin{equation}
{\bf M}_\nu^{eff}=
-m^{PSS}\,({\mathrm{M}}_{\chi^0}^{PSS})^{-1}\,(m^{PSS})^T=
\frac{M_1 \tilde g^2_d + M_2 \tilde {g'}^2_d}{4\det{M_{\chi^0}^{PSS}}}
\left[\begin{array}{cccc}
\Lambda_1^2        & \Lambda_1\Lambda_2 & \Lambda_1\Lambda_3 \cr
\Lambda_2\Lambda_1 & \Lambda_2^2        & \Lambda_2\Lambda_3 \cr
\Lambda_3\Lambda_1 & \Lambda_3\Lambda_2 & \Lambda_3^2
\end{array}\right],
\label{treenumass2}
\end{equation}
with $\Lambda_i=\mu b_i v_u + \epsilon_i v_d$, and with the determinant 
of the neutralino submatrix equal to,
\begin{equation}
\det{M_{\chi^0}^{PSS}}=-\mu^2 M_1 M_2 + \frac{1}{2} v_uv_d\mu \left( 
M_1 \tilde g_u\tilde g_d + M_2 \tilde g'_u \tilde g'_d \right)
+\textstyle{\frac{1}{16}} v_u^2v_d^2 
\left(\tilde g'_u \tilde g_d - \tilde g_u \tilde g'_d \right)^2.
\label{detNeut2}
\end{equation}
As it is well known, the neutrino mass matrix in eq.~(\ref{treenumass2})
has only one eigenvalue different from zero, and quantum corrections must 
be added in order to generate a solar mass.

%%%%%%%%%%%%%%%%%%%%%%%%%%%%%%%%%%%%%%%%%%%%%%%%%%%%%%%%%%%%%
\section{Higgs Decoupling Limit}
%%%%%%%%%%%%%%%%%%%%%%%%%%%%%%%%%%%%%%%%%%%%%%%%%%%%%%%%%%%%%

The Higgs decoupling limit in the MSSM, or in the non-supersymmetric two 
Higgs doublet model 2HDM, is the regime where the lightest Higgs scalar 
mass is much smaller that all the other Higgs boson masses 
\cite{Gunion:2002zf}. It has been studied in detail because if realized
in nature, it will be experimentally difficult to distinguish the lithest 
Higgs boson properties from the ones of the SM Higgs boson.

If we neglect the running of the Higgs potential parameters, in first
approximation the Higgs sector in PSS is analogous to the one in the 
MSSM. At tree level we have for the neutral CP-even Higgs bosons the
following masses,
\begin{equation}
m^2_{h,H}=\frac{1}{2}\left(m_A^2+m_Z^2\right)\mp\frac{1}{2}\sqrt{
\left(m_A^2+m_Z^2\right)^2-4m_A^2m_Z^2c^2_{2\beta}}
\end{equation}
as a function of the neutral CP-odd Higgs mass $m_A$ and $\tan\beta$.
Although $m_h$ receive large quantum corrections, they are negligible 
for the heavy Higgs H when $m_A$ is large. Thus the tree-level formula for
$m_H$ is adequate in the decoupling limit. In this limit, when 
$m_A\gg m_Z$, we have
\begin{equation}
m_H^2\approx m_A^2+m_Z^2\sin^2(2\beta)
\end{equation}
which is a very good approximation already for $m_A>200$ GeV. The 
CP-even Higgs mass matrix is diagonalized by a rotation with an angle 
$\alpha$, which satisfies at tree-level,
\begin{equation}
\cos^2(\alpha-\beta)=\frac{m_h^2(m_Z^2-m_h^2)}{m_A^2(m_H^2-m_h^2)}.
\end{equation}
The quantity $\cos(\alpha-\beta)$ is important because it is equal
to the ratio between the heavy Higgs coupling to two $Z$ bosons 
(and two $W$ bosons) and the same couplings for the SM Higgs boson,
$H_{SM}$. Thus, it is a key parameter for the determination of the
$H$ production cross section in association with gauge bosons. 
Conversely, $\sin(\alpha-\beta)$ is proportional to the light Higgs 
boson couplings to two gauge bosons. In this limit 
we have,
\begin{equation}
\cos^2(\alpha-\beta) \approx \frac{m_Z^4\sin^2(4\beta)}{4m_A^4} \quad.
\label{cosApp}
\end{equation}
This means
that $H$ decouples from the low energy theory. Thus the $h$ 
production cross section becomes ever more similar to the $H_{SM}$ one. 
The same happens to the $h$ couplings to fermions: they become similar 
to the $H_{SM}$ couplings in the decoupling limit, making it an 
experimental challenge to differentiate a $H_{SM}$ from a
$h$ in this scenario.

%%%%%%%%%%%%%%%%%%%%%%%%%%%%%%%%%%%%%%%%%%%%%%%%%%%%%%%%%%%%%
\section{Loop Corrections to Neutrino Masses
in the Decoupling Limit}
%%%%%%%%%%%%%%%%%%%%%%%%%%%%%%%%%%%%%%%%%%%%%%%%%%%%%%%%%%%%%

In PSS the only loops capable to contribute to the solar mass are
loops involving neutralinos and neutral Higgs bosons. 
\begin{center}
\vspace{-50pt} \hfill \\
\begin{picture}(200,120)(0,23) % y_2 controls equation position
%
% Top left graph
%
\ArrowLine(20,50)(80,50)
\Text(50,60)[]{$\nu_j$}
\ArrowArcn(110,50)(30,180,0)
\Text(110,13)[]{$h,H,A$}
\DashCArc(110,50)(30,180,0){3}
\Text(110,95)[]{$\chi^0_k$}
\ArrowLine(140,50)(200,50)
\Text(170,60)[]{$\nu_i$}
\end{picture}
\vspace{30pt} \hfill \\
\end{center}
\vspace{-10pt}
Radiative corrections to the effective neutrino mass matrix are of the form,
\begin{equation}
\Delta\Pi_{ij}=A\Lambda_i\Lambda_j+
B(\Lambda_i\epsilon_j+\Lambda_j\epsilon_i)+
C\epsilon_i\epsilon_j,
\label{DpiH2HDM}
\end{equation}
where the $A$-term is the only one that receives tree-level contributions,
as indicated in eq.~(\ref{treenumass2}). Charged and neutral gauge bosons
contribute only to the $A$-term, {\it i.e.}, to the atmospheric mass. The 
charged Higgs boson contributes to $A$ and $B$-terms. However, the 
$B$-term is scale dependent and can be rendered zero with an appropriate 
choice for the arbitrary substraction scale $Q$. Therefore, charged Higgs bosons
do not contribute neither to the solar mass \cite{Diaz:2006ee}. 
This leaves only the neutral 
Higgs bosons, whose contribution to the $C$ term are,
\begin{eqnarray}
\Delta\Pi_{ij}^{h,H}\Big|_{\epsilon\epsilon}&=&-\frac{1}{16\pi^2}\sum_{k=1}^4
\big(F_k^{h,H}\big)^2
\epsilon_i\epsilon_j
m_{\chi_k^0}B_0(0;m_{\chi_k^0}^2,m_{h,H}^2)\,,
\nonumber\\
\Delta\Pi_{ij}^{A}\Big|_{\epsilon\epsilon}&=&\frac{1}{16\pi^2}\sum_{k=1}^4
\big(F_k^A\big)^2
\epsilon_i\epsilon_j
m_{\chi_k^0}B_0(0;m_{\chi_k^0}^2,m_A^2)\,,
\label{3loops}
\end{eqnarray}
where the sum is over all four neutralinos, with masses $m_{\chi_k^0}$,
and $B_0$ is the usual Veltman function for two-point Green functions. Note
that $h$ and $H$ contributions have an overall minus sign, while the $A$ 
contribution has not. The reason is that $A$ has CP-odd couplings to 
fermions. The couplings $F_k$ are equal to,
\begin{eqnarray}
F_k^h&=&\frac{\cos(\alpha-\beta)}{2\mu s_\beta}
\left(g N^*_{k2}-g' N^*_{k1}\right),
\nonumber\\
F_k^H&=&\frac{\sin(\alpha-\beta)}{2\mu s_\beta}
\left(g N^*_{k2}-g' N^*_{k1}\right),
\label{EandF2}\\ \nonumber
F_k^A&=&\frac{1}{2\mu s_\beta}
\left(g N^*_{k2}-g' N^*_{k1}\right).
\end{eqnarray}
The factor in the parenthesis indicates that it is the gaugino component 
of the neutralinos the one that contributes to the solar neutrino mass. 
In addition, the $B_0$ Veltman's function for zero external momentum is, 
\begin{equation}
B_0(0;M^2,m^2)=\Delta+1-\frac
{M^2\ln{\frac{M^2}{Q^2}}-m^2\ln{\frac{m^2}{Q^2}}}
{M^2-m^2}
\label{B0}
\end{equation}
where $Q$ is the arbitrary substraction scale and $\Delta$ is the regulator. 
In this way, the one-loop contribution to the $C$ coefficient of the 
$\epsilon_i\epsilon_j$ term in eq.~(\ref{DpiH2HDM}) is,
\begin{eqnarray}
C^{AHh}&=&\frac{1}{64\pi^2\mu^2s_\beta^2}\sum_{k=1}^4m_{\chi_k^0}
\left(g N_{k2}-g' N_{k1}\right)^2
\label{ChHA}\\\nonumber &&
\left[ B_0(0;m_{\chi_k^0}^2,m_A^2) -
\sin^2(\alpha-\beta)B_0(0;m_{\chi_k^0}^2,m_H^2) -
\cos^2(\alpha-\beta)B_0(0;m_{\chi_k^0}^2,m_h^2) \right]\,.
\end{eqnarray}
The fact that the divergent term in $B_0$ is independent of all masses 
implies that the $C$ coefficient is finite, which in turn leads to the 
finiteness of the solar mass.

The neutrino masses generated from eq.~(\ref{DpiH2HDM}) include a 
massless neutrino $m_{\nu_1}=0$, and the two massive ones given by,
\begin{equation}
m_{\nu_{3,2}}=\frac{1}{2}\left( A|\vec\Lambda|^2+
C|\vec\epsilon\,|^2\right) \pm \frac{1}{2}\sqrt{
\left( A|\vec\Lambda|^2+C|\vec\epsilon\,|^2\right)^2-
4AC|\vec\Lambda\times\vec\epsilon\,|^2}
\end{equation}
where the sign is chosen such that $|m_{\nu_{2}}|<|m_{\nu_{3}}|$.
We are interested in the behavior of the solar mass in the Higgs 
decoupling limit, where $m_A\gg m_Z$. The $B_0$ Veltman function 
in eq.~(\ref{B0}) has the following expansion when one of the masses
is much larger than the other, $M\gg m$,
\begin{equation}
B_0(0;M^2,m^2)\approx \Delta-\ln\frac{M^2}{Q^2}+1-
\frac{m^2}{M^2}\ln\frac{M^2}{m^2}-\frac{m^4}{M^4}\ln\frac{M^2}{m^2}
\end{equation}
In the regime where the neutralinos are much lighter than the CP-odd
Higgs mass, we find for the $C$ coefficient,
\begin{equation}
C^{AHh} \approx \frac{m_Z^2\sin^22\beta}
{64\pi^2\mu^2s_\beta^2m_A^2}
\sum_{k=1}^4m_{\chi_k^0}
\left(g N_{k2}-g' N_{k1}\right)^2
\end{equation}
where the term in parenthesis correspond to the zino component of each
neutralino. This motivates us to define,
\begin{equation}
\langle m_{\widetilde Z} \rangle \equiv
\sum_{k=1}^4m_{\chi_k^0}
\left(c_W N_{k2}-s_W N_{k1}\right)^2
\label{zinoM}
\end{equation}
as the zino effective mass. In this way, we find that the CP-odd Higgs 
mass is related to the solar neutrino mass difference by the following 
simple relation,
\begin{equation}
m_A^2 \approx
\frac{g^4\, m_Z^2\cos^2\beta}{64\pi^2 c_W^4}
\frac{\langle m_{\widetilde Z} \rangle m_{\widetilde\gamma}}{M_1M_2}
\sqrt{\frac{\delta}{1+\delta}}
\frac{|\vec\Lambda\times\vec\epsilon\,|^2}
{\mu^4 \Delta m^2_{sol}}
\label{AppmA}
\end{equation}
where 
$
\delta=\Delta m^2_{sol}/\Delta m^2_{atm}\approx 0.035
$
and $m_{\widetilde\gamma}=c_W^2M_1+s_W^2M_2$ is the photino mass.

This formula is remarkable. First we notice that the CP-odd Higgs mass 
squared is inversely proportional to the solar mass difference. The 
reason behind this feature is as follows. In the Higgs decoupling limit 
the light Higgs has SM-like couplings and does not contribute to the 
solar mass. More precisely, $h$ contribution to the neutrino mass matrix 
is proportional to $\cos^2(\alpha-\beta)$, which rapidly approaches zero 
as $m_A\gg m_Z$, as indicated by eq.~(\ref{cosApp}). The other two neutral 
Higgs bosons, $H$ and $A$, have large contributions to the neutrino mass 
matrix, but with opposite signs, and as it can be seen already
from eq.~(\ref{3loops}) they tend to cancel each other in the 
decoupling limit. Therefore, the fact that Supersymmetry forces 
$m_H\rightarrow m_A$ when $m_A\gg m_Z$, produces a fine cancellation that 
eventually generates a small solar mass. This fine cancellation is not a 
fine-tuning because it is a cancellation forced by symmetry. In addition,
the CP-odd Higgs mass is dependent on the atmospheric mass through the
ratio $\delta$ between solar at atmospheric scales. Thus, the atmospheric 
mass also affects $m_A^2$ inversely although in an indirect way. Our
model explains the smallness of $\delta$ because the atmospheric mass
is generated at tree-level while the solar mass is generated at one-loop.

The CP-odd squared mass is proportional to 
$|\vec\Lambda\times\vec\epsilon\,|$, with an extra term $\mu^4$ in the 
denominator. This cross product 
is there because if $\vec\epsilon$ and $\vec\Lambda$ are parallel, the 
symmetry of the neutrino mass matrix observed at tree-level in 
eq.~(\ref{detNeut2}) is not removed, and no solar mass is generated. 
Furthermore, $m_A^2$ is proportional to the $Z$ boson mass, 
indicating that a Majorana neutrino mass needs not only R-Parity 
violation but a  broken $SU(2)_L$ symmetry as well.
Additionally, $m_A^2$ is proportional to $\cos^2\beta$
because the relevant vacuum expectation value is $\langle H_d \rangle$.

We also notice that the CP-odd Higgs mass squared is proportional to the 
effective zino mass and to the photino mass, normalized by both relevant 
gaugino masses. The appearance of the photino mass is due to the tree-level
contribution to the effective neutrino mass matrix. The origin of the zino 
mass is that it is the zino component of each neutralino the one that 
contributes to the neutrino mass matrix, and it multiplies the neutralino 
masses from the fermionic propagator. In addition, for this mechanism to 
work, R-Parity must be broken, and the neutrino mixing with down higgsinos 
in the one hand, and Higgs bosons mixing with sneutrinos in the other hand, 
are crucial. The following schematic diagram, corresponding to the heavy 
Higgs boson loops, may help to understand the origin of the one-loop 
contributions to the neutrino mass matrix,
\begin{center}
\vspace{-50pt} \hfill \\
\begin{picture}(200,120)(0,23) % y_2 controls equation position
%
% Top left graph
%
\ArrowLine(0,50)(40,50)
\Text(20,60)[]{$\nu_j$}
\Line(40,50)(80,50)
\Text(60,59)[]{$\widetilde H_d$}
\BCirc(40,50){3}
\Text(40,10)[]{$\epsilon_j/\mu$}
\LongArrow(40,15)(40,45)
\DashCArc(110,50)(30,180,0){3}
\Text(110,13)[]{$H$}
\Text(77,33)[]{$H_d$}
\GCirc(125,25){3}{0}
\Text(133,17)[]{$c_\alpha$}
\GCirc(95,25){3}{0}
\Text(90,17)[]{$c_\alpha$}
\Text(145,33)[]{$H_d$}
\CArc(110,50)(30,0,180)
\Text(79,70)[]{$\widetilde Z$}
\Text(110,90)[]{$\chi^0_k$}
\GCirc(95,75){3}{0}
\Text(38,94)[]{$c_WN_{k2}-s_WN_{k1}$}
\LongArrow(79,91)(92,79)
\GCirc(125,75){3}{0}
\Text(184,94)[]{$c_WN_{k2}-s_WN_{k1}$}
\LongArrow(143,91)(129,79)
\Text(143,70)[]{$\widetilde Z$}
\Line(140,50)(180,50)
\Text(160,59)[]{$\widetilde H_d$}
\ArrowLine(180,50)(220,50)
\Text(200,60)[]{$\nu_i$}
\BCirc(180,50){3}
\Text(180,10)[]{$\epsilon_i/\mu$}

\LongArrow(180,15)(180,45)
\end{picture}
\vspace{30pt} \hfill \\
\end{center}
\vspace{10pt}
The supersymmetric vertex behind this diagram is the zino coupling to 
down-Higgs and down-higgsino. For this reason, the zino component from
each neutralino is selected, $c_WN_{k2}-s_WN_{k1}$, which weights the 
corresponding neutralino mass picked up from the propagator. In addition, 
the down-Higgs component from the heavy Higgs $H$, given by $c_\alpha$, 
is selected. These mixings are represented in the diagram by full 
circles, as oppose to open circles which violate R-Parity. Indeed,
R-Parity is violated at the mixing between neutrinos and down-higgsinos.
These two mixings at the external legs also violate lepton number by 
two units, as it should be for a Majorana neutrino mass.

The second diagram is,
\begin{center}
\vspace{-50pt} \hfill \\
\begin{picture}(200,120)(0,23) % y_2 controls equation position
%
% Top left graph
%
\Text(40,58)[]{$\nu_j$}
\ArrowLine(40,50)(80,50)
\Text(50,10)[]{$\epsilon_j s_\alpha/\mu t_\beta$}
\LongArrow(70,13)(91,23)
\DashCArc(110,50)(30,180,0){3}
\Text(110,13)[]{$H$}
\Text(77,33)[]{$\widetilde\nu_j$}
\BCirc(125,25){3}
\BCirc(95,25){3}
\Text(145,33)[]{$\widetilde\nu_i$}
\CArc(110,50)(30,0,180)
\Text(79,70)[]{$\widetilde Z$}
\Text(110,90)[]{$\chi^0_k$}
\GCirc(95,75){3}{0}
\Text(38,94)[]{$c_WN_{k2}-s_WN_{k1}$}
\LongArrow(79,91)(92,79)
\GCirc(125,75){3}{0}
\Text(184,94)[]{$c_WN_{k2}-s_WN_{k1}$}
\LongArrow(143,91)(129,79)
\Text(143,70)[]{$\widetilde Z$}
\ArrowLine(140,50)(180,50)
\Text(180,58)[]{$\nu_i$}
\Text(170,10)[]{$\epsilon_i s_\alpha/\mu t_\beta$}
\LongArrow(149,13)(129,23)
\end{picture}
\vspace{30pt} \hfill \\
\end{center}
\vspace{10pt}
where the supersymmetric vertex supporting the diagram is the zino coupling
to a neutrino and a sneutrino. The presence of the zino component of each 
neutralino is explained by the same argument as before. But in this case, 
R-Parity and lepton number violation appear in the Higgs boson mixing with
sneutrinos. The magnitude of this mixing, indicated in the diagram, is 
explained in ref.~\cite{Diaz:2006ee}. Diagrams with one of each supersymmetric 
vertices also contribute, but are not shown.

%%%%%%%%%%%%%%%%%%%%%%%%%%%%%%%%%%%%%%
\section{Numerical Results}
%%%%%%%%%%%%%%%%%%%%%%%%%%%%%%%%%%%%%%

In our numerical analysis the contributions to the neutrino
mass matrix in eq.~(\ref{DpiH2HDM}) were calculated and their influence
on the neutrino observables such as mass differences
and mixing angles was studied.
The agreement with the experimental boundaries ($3\sigma$) \cite{Maltoni:2004ei}
was quantified by calculating
\begin{equation}\label{xi2}
\chi^2=\left(\frac{10^3\Delta m^2_{atm}-2.4}{0.4}\right)^2+
\left(\frac{10^5\Delta m^2_{sol}-7.7}{0.6}\right)^2+
\left(\frac{\sin^2\theta_{atm}-0.505}{0.165}\right)^2+
\left(\frac{\sin^2\theta_{sol}-0.33}{0.07}\right)^2.
\end{equation}
Additionally it was demanded that the upper bounds 
$\sin^2 \theta_{reac}<0.05$ and $m_{\beta \beta}<0.84$ eV
have to be fulfilled.
Thus, the model is in agreement with the experimental
values if $\chi^2<4$.
As a working scenario, we select a typical point in the PSS parameter 
space, denoted P$_t$. This scenario consists of fixed values for the
gaugino and higgsino mass parameters, $\tan\beta$, the light CP-even and
CP-odd Higgs masses, and the BRpV parameters, all given in table \ref{tab1}.
\begin{table}
\caption{PSS and RpV parameters and neutrino observables for the working
scenario P$_t$ \label{tab1}}
\bigskip
\begin{minipage}[h]{7 cm}
\begin{tabular}{cccc}
\hline
Susy-Parameter & P$_t$ & Scanned range & Units \\
\hline \hline
$\tan\beta$ & 10 & [2,50] & -  \\
$|\mu|$ & 450 & [0,1000] & GeV  \\
$M_2$ & 300 & [80,1000] & GeV  \\
$M_1$  & 150 & $M_2/2$ & GeV  \\
$m_h$  & 120 & [114,140] & GeV  \\
$m_A$  & 1000 & [50,6000] & GeV  \\
$Q$  & 951.7 & - &GeV  \\
\hline
RpV-Parameter\\
\hline 
$\epsilon_1$ & 0.0346 & - &GeV  \\
$\epsilon_2$ & 0.2516 & [-1,1] &GeV  \\
$\epsilon_3$ & 0.3504 & [-1,1] &GeV  \\
$\Lambda_1$  & -0.0259 & [-1,1] & GeV${}^2$  \\
$\Lambda_2$  & -0.0011 & - & GeV${}^2$  \\
$\Lambda_3$  & 0.0709 & [-1,1] &GeV${}^2$  \\
\hline \hline \label{tab:MSSMabc}
\end{tabular}
\end{minipage}
\begin{minipage}[t]{7 cm}
\begin{tabular}{ccc}
\hline
Observable & Solution & Units \\
\hline \hline
$\Delta m_{atm}^2$ & 2.45$\times10^{-3}$ & eV${}^2$ \\
$\Delta m_{sol}^2$ & 7.9$\times10^{-5}$ & eV${}^2$ \\
$\tan^2\theta_{atm}$ & 0.824 & - \\
$\tan^2\theta_{sol}$ & 0.487 & - \\
$\tan^2\theta_{13}$ & 0.027 & - \\
$m_{ee}$ & 0.0016 & eV \\
\hline \hline
\\ \\ \\ \\ \\ \\ \\ \\
 \label{tab:sol}
\end{tabular}
\end{minipage}
\end{table}
This scenario satisfies the neutrino experimental constraints, predicting
atmospheric and solar mass differences and mixing angles well within 
the $3\sigma$ regions indicated in eq.~(\ref{xi2}). In addition, the
predicted reactor angle and the neutrinoless double-beta decay mass 
parameter are below the upper bound. All these prediction of our P$_t$
scenario are given in table \ref{tab1}.

Next we scan the parameter space varying the PSS parameters according 
to the intervals indicated in table \ref{tab1}.
%
%%%%%%%%%% FIGURE %%%%%%%%%%%%%%%%%%
\begin{figure}[htbp]
  \centering
  \begin{minipage}[b]{7 cm}
\epsfig{file=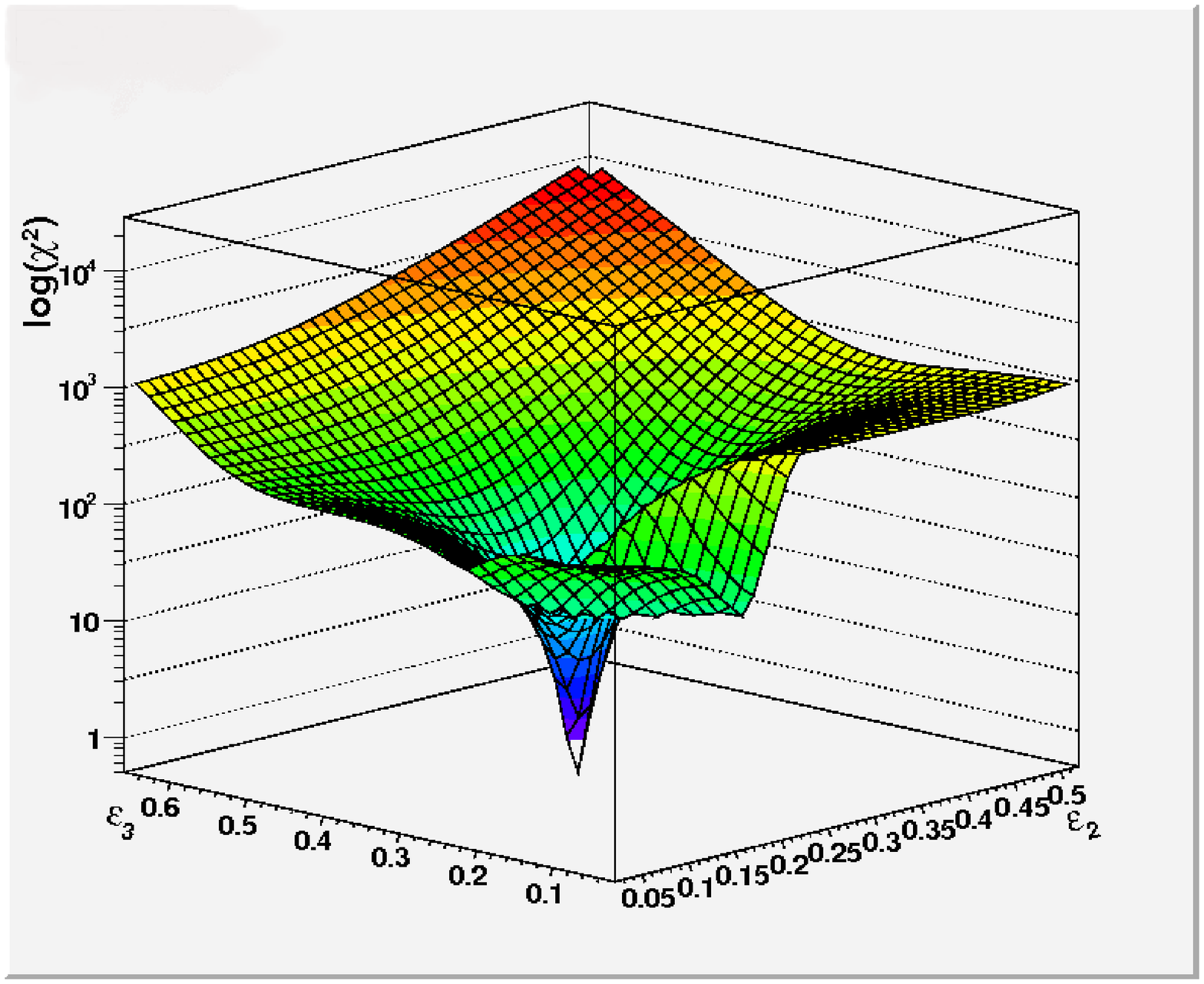,
width=1.0\textwidth}
\caption{\it $\chi^2$ in dependence of $\epsilon_2$ and $\epsilon_3$, while
the other parameters are fixed around the central value
from table \ref{tab1}.}
\label{eps2eps3}
  \end{minipage}
\hspace{0.5 cm}
  \begin{minipage}[b]{7 cm}
\centerline{\protect\vbox{\epsfig{file=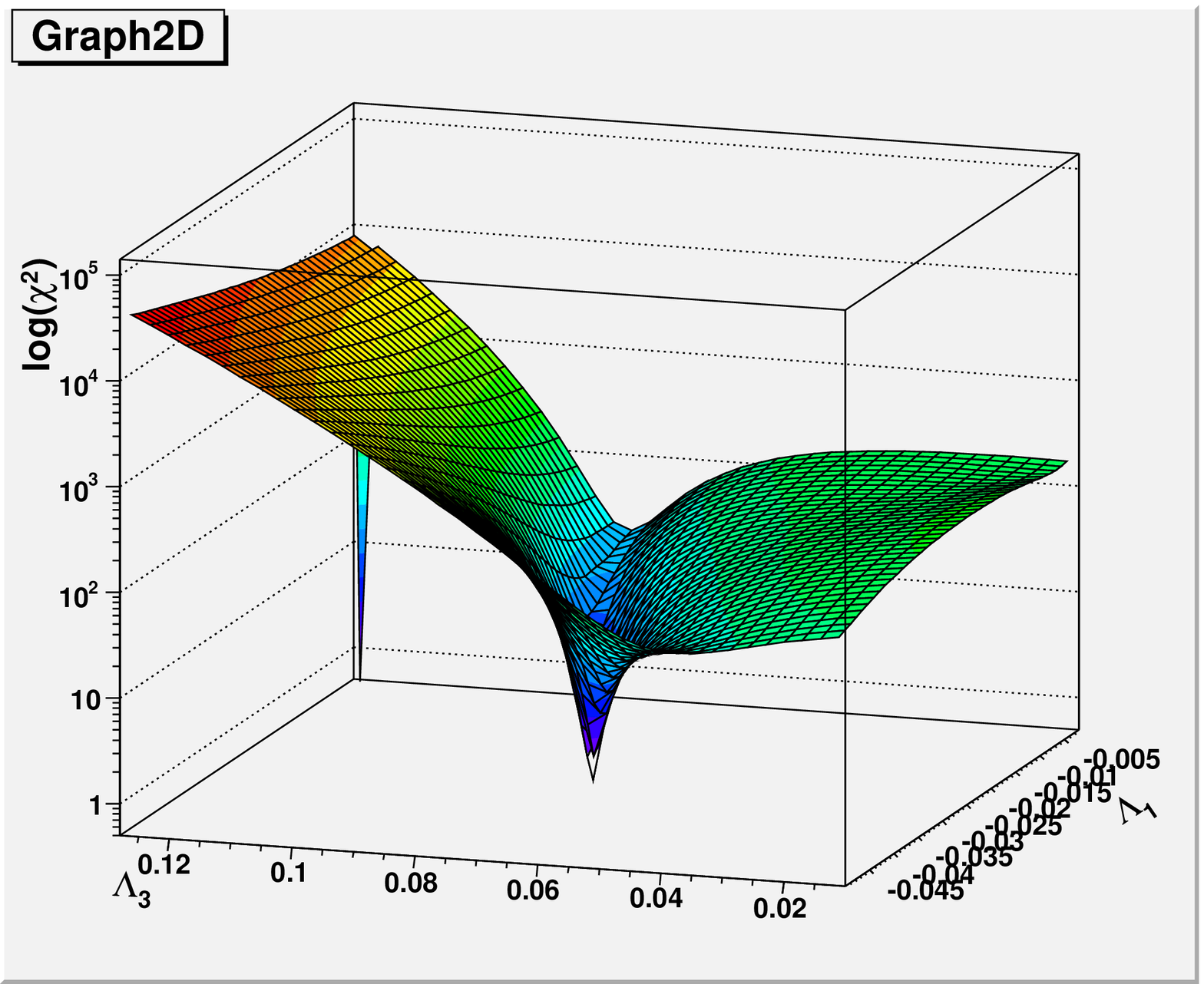,
width=1.0\textwidth}}}
\caption{\it $\chi^2$ in dependence of $\Lambda_1$ and $\Lambda_3$,
while the other parameters are fixed around the central value from table
\ref{tab1}.}
\label{lam3lam1}
  \end{minipage}
\end{figure}
%%%%%%%%%%%%%%%%%%%%%%%%%%%%%%%%%%%%%%5
%
In Fig.~\ref{eps2eps3}, we vary $\epsilon_2$ and $\epsilon_3$ keeping all the
other parameters as in P$_t$, and plot the logarithm of $\chi^2$ associated to 
each point in parameter space. Solutions with $\chi^2<4$ are clearly visible, 
they are compatible with experiments, and P$_t$ is inside this 
region. The value for $\chi^2$ grows fast as we deviate from the experimentally 
accepted region around P$_t$. In Fig.~\ref{lam3lam1} we see the dependence on 
$\Lambda_1$ and $\Lambda_3$ while keeping the rest of the PSS parameters as
indicated by P$_t$. Again there is a steep growth on $\chi^2$ when we deviate
from the experimentally accepted region. Of course, these experimentally 
compatible regions move around in the $\epsilon_2$-$\epsilon_3$ and 
$\Lambda_1$-$\Lambda_3$ planes when we change the fixed values of the other PSS 
parameters.

%
%%%%%%%%%% FIGURE %%%%%%%%%%%%%%%%%%
\begin{figure}[htbp]
  \centering
  \begin{minipage}[b]{7 cm}
\epsfig{file=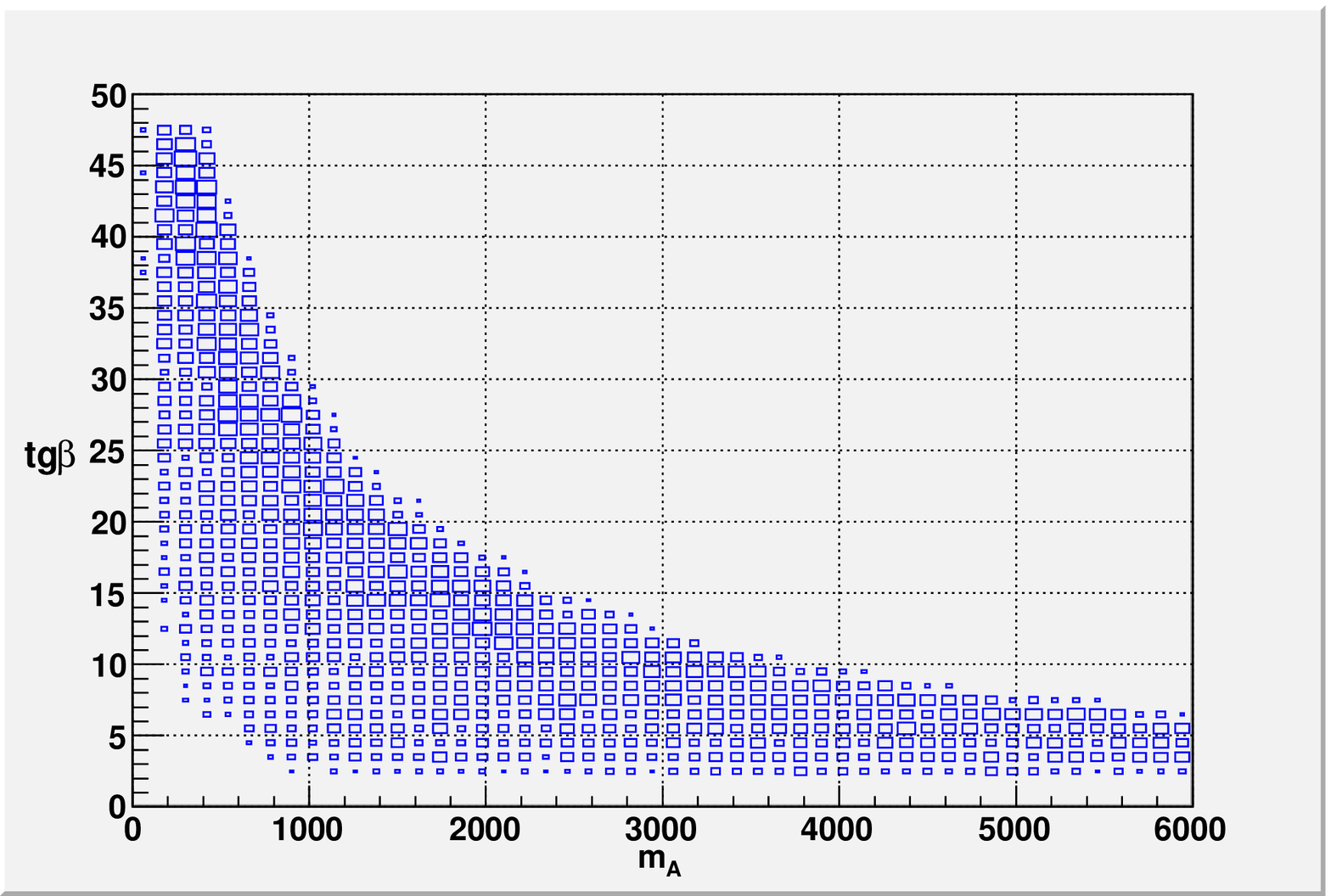,
width=1.0\textwidth}
\caption{\it Relation between $\tan\beta$ and $m_A$,
while varying the other parameters as indicated in \ref{tab1}.}
\label{tbma}
  \end{minipage}
\hspace{0.5 cm}
  \begin{minipage}[b]{7 cm}
\centerline{\protect\vbox{\epsfig{file=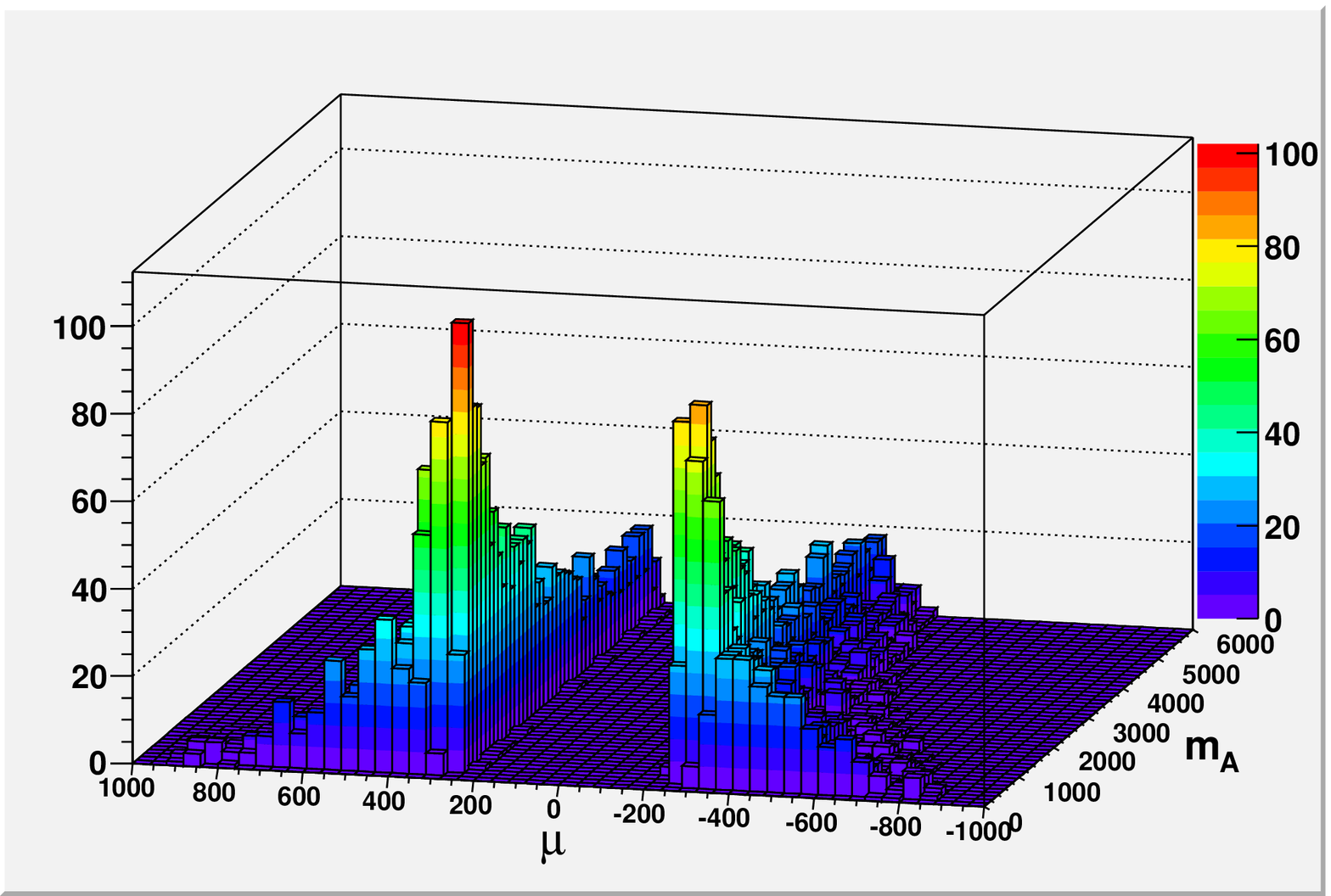,
width=1.0\textwidth}}}
\caption{\it Frequency plot in the $\mu$-$m_A$ plane,
for points in parameter space that are consistent with neutrino experiments.}
\label{muma}
  \end{minipage}
\end{figure}
%%%%%%%%%%%%%%%%%%%%%%%%%%%%%%%%%%%%%%5
%
For the scan shown in Figs.~\ref{tbma} and \ref{muma} we keep the BRpV parameters
$\epsilon_i$ and $\Lambda_i$ fixed to 
their P$_t$ values, and vary the rest of the
PSS parameters as indicated in table \ref{tab1}. In Fig.~\ref{tbma} we have the 
relation between $\tan\beta$ and the CP-odd Higgs mass $m_A$. The scan shows that 
an agreement with the neutrino observables is disfavored if both susy parameters 
$\tan\beta$ and $m_A$ take simultaneously large values. Due to this correlation
a measurement of the value of one of the two parameters might give important 
information on the value of the other, even if the rest of the susy parameters 
are not known. A detailed study shows that 
the excluded region in the $(\tan \beta-m_A)$ plane comes
due to the observed constraint on $\tan^2 \theta_{atm}$.
In Fig.~\ref{muma} we show the relation between the higgsino mass parameter 
$\mu$ and $m_A$ from the same scan as before. This reveals that for large 
$m_A\approx6000$~GeV good results compatible with neutrino experiments are only 
obtained for $200<|\mu|< 350$~GeV, whereas for smaller $m_A\approx 1000$~GeV
the preferred region for $\mu$ widens up to $200<|\mu|< 600$~GeV. A detailed 
study shows that the observed values for 
$\Delta m^2_{sol}$ and $\Delta m^2_{atm}$ forbid
solutions with $|\mu|<200$~GeV.
%
%%%%%%%%%% FIGURE %%%%%%%%%%%%%%%%%%
\begin{figure}
\centerline{\protect\vbox{\epsfig{file=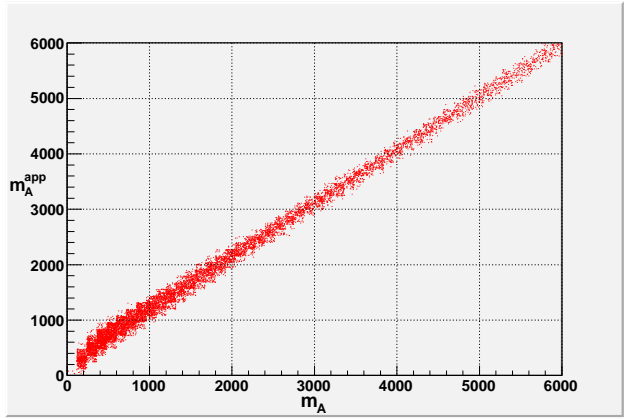,width=0.5\textwidth}}}
\caption{\it Approximated $m_A^{app}$ as a function of 
its exact value $m_A$, for
a scan with $10^4$ random points.}
\label{maaprox}
\end{figure}
%%%%%%%%%%%%%%%%%%%%%%%%%%%%%%%%%%%%
%
Finally, in order to test the accuracy of the approximated formula given in 
eq.~(\ref{AppmA}), we perform a scan with $10^4$ random points.
In Fig.~\ref{maaprox} we include the points 
that satisfy the neutrino experimental constraints, and we see that the 
approximated formula works in a very large range of $m_A$ values, with larger 
percentage errors for smaller $m_A$, which is expected.

%%%%%%%%%%%%%%%%%%%%%%%%%%%%%%%%%%%%%%%%%%%%%%%%%%%%%%%%%%%%%%%%%%%%%%%%%%%%%%%
\section{Summary}
%%%%%%%%%%%%%%%%%%%%%%%%%%%%%%%%%%%%%%%%%%%%%%%%%%%%%%%%%%%%%%%%%%%%%%%%%%%%%%%
\label{conclusions}
This paper explored Partial Split Supersymmetry with RpV as a model for neutrino 
masses. It was shown that at the one-loop level the model can give predictions 
that are in very good agreement with all present experimental values for the 
neutrino sector. In contrast to this good agreement in PSS, it is not possible 
to generate all the neutrino mass parameters correctly in standard Split 
Supersymmetry with RpV. The difference between both models, lies solely in the 
fact that PSS allows for a larger Higgs sector, which contains the mass 
eigenstates $A$ and $H$ in addition the standard model like state $h$. A 
continuous transition from PSS to SS can be achieved by raising the values for 
the heavy Higgs masses $m_A$ and $m_H$ (Higgs decoupling limit). An analytical 
study of this limit ($m_A\gg m_h$) reveals an approximate formula for PSS in 
which a large value for $m_A$ is directly connected to a small solar neutrino 
mass difference $\Delta m_{sol}^2$. Therefore, the small observed value for 
$\Delta m_{sol}^2$ favors large values of $m_A$ up to 6~TeV. Such large values 
for $m_A$ would make PSS virtually indistinguishable from SS by using any 
observable other than neutrino masses.

%%%%%%%%%%%%%%%%%%%%%%%%%%%%%%%%%%%%%%%%%%%%%%%%%%%%%%%%%%%%%%%%%%
{\bf Note Added:}
{\small 
While this article was being written, we read the paper ``SUSY Splits, But 
Then Returns'' (arXiv:0909.5430) by Prof.~Raman Sundrum \cite{Sundrum:2009gv}, 
where models with two light Higgs doubles are also referred to as Partial 
Split Supersymmetry.
}
%%%%%%%%%%%%%%%%%%%%%%%%%%%%%%%%%%%%%%%%%%%%%%%%%%%%%%%%%%%%%%%%%

%%%%%%%%%%%%%%%%%%%%%%%%%%%%%%%%%%%%%%%%%%%%%%%%%%%%%%%%%%%%%%%%%%
\begin{acknowledgments}
{\small 
The work of M.A.D. was partly funded by Conicyt-PBCT grant 
No.~ACT028 (Anillo Centro de Estudios Subat\'omicos), and by 
Conicyt-PBCT grant ACI35. B.K. was funded by Conicyt-PBCT grant PSD73.}
\end{acknowledgments}
%%%%%%%%%%%%%%%%%%%%%%%%%%%%%%%%%%%%%%%%%%%%%%%%%%%%%%%%%%%%%%%%%

%%%%%%%%%%%%%%%%%%%%%%%%%%%%%%%%%%%%%%%%%%%%%%%%%%%%%%%%%%%%%%%%%%%%%%%%%%%%%%

\end{document}